\documentclass[%
 reprint,
 amsmath,amssymb,
 aps,
]{revtex4-2}

\usepackage{graphicx}
\usepackage{dcolumn}
\usepackage{bm}
\usepackage{placeins}
\usepackage{stackengine}
\usepackage{color}
\usepackage{hyperref}

\newcommand{\eff}{{e\!f\!\!f}}

\begin{document}

\preprint{APS/123-QED}

\title{Ray-tracing laser-deposition model for 
plasma particle-in-cell simulation}

\author{A. Hyder}
 \email{a.hyder@columbia.edu}
\affiliation{
 Dept. of Applied Physics and Applied Mathematics, Columbia University, New York, NY 10027, USA
}

\author{W. Fox}
\email{wfox@pppl.gov}
\affiliation{
 Dept. of Astrophysical Sciences, Princeton University, Princeton, NJ 08544, USA
}%
\affiliation{
 Princeton Plasma Physics Laboratory, Princeton, NJ 08540, USA
}%
\author{K. V. Lezhnin}
\affiliation{
 Princeton Plasma Physics Laboratory, Princeton, NJ 08540, USA
}%
\author{S. R. Totorica}
\affiliation{
 Dept. of Astrophysical Sciences, Princeton University, Princeton, NJ 08544, USA
}
\affiliation{Dept. of Astro-fusion Plasma Physics (AFP), Headquarters for Co-Creation Strategy, National Institutes of Natural Sciences, Tokyo 105-0001, Japan}

\date{\today}

\begin{abstract}
We develop a ray-tracing model for laser-plasma interaction suitable for coupling in-line into 
kinetic particle-in-cell plasma simulation.  The model is
based on inverse Bremsstrahlung absorption and includes
oblique incidence effects and reflection at the
critical surface.  The energy deposition is 
given to electrons by randomized kicks to momentum.
The model is verified 
against analytic solutions and a 2-D laser
ray-tracing code.
\end{abstract}

\maketitle


\section{\label{sec:introduction}Introduction}

Kinetic plasma simulations are a valuable tool for understanding high-energy-density (HED) plasma systems driven by intense radiation and energy sources \cite{Rinderknecht_2018}. Modern high-energy lasers, such as those at the National Ignition Facility (NIF) and OMEGA facility, couple several kJ of energy into $\sim$mm volumes of plasma within ns-timescales \cite{Danson_2019}. These lasers readily heat plasmas to keV temperatures, producing plasmas with long mean free paths.  Furthermore, these plasmas can self-generate magnetic fields of 10's of Tesla sufficient to magnetize the plasma.  Kinetic simulations can be valuable tools for understanding these systems, as they do not assume local Maxwellian distributions in the plasma. They can more readily handle kinetic effects, such as particle counterstreaming and temperature anisotropy, as well as simulate kinetic plasma processes, such as the formation of collisionless shocks \cite{Fox_2018, Schaeffer_2020, Lezhnin_2021}, magnetic field generation by Weibel instability \cite{Fox_2013}, and magnetic reconnection \cite{Fox_2011, Matteucci_2018}. To improve fidelity when designing and interpreting these experiments via simulations, a key step is to accurately simulate how the laser propagates and heats the plasma. For the laser pulses of interest in the present work, we consider how to model laser deposition via inverse Bremsstrahlung (IB), which is the collisional absorption of the laser photons by plasma.

In this paper, we develop a radiation transport model for laser propagation and absorption suitable for coupling in-line into a kinetic plasma particle-in-cell simulation. The model traces laser intensity along rays using geometric optics, with laser absorption and energy deposition based on inverse Bremsstrahlung (IB) \cite{Dawson_1964}. 
In this class of model, the laser-intensity envelope is simulated rather than the full-wave evolution of the electromagnetic (EM) fields. This relaxes the simulation requirements so that the laser wavelength and frequency do not have to be resolved by the simulation. This approach is often used in radiation hydrodynamics simulation (e.g., Refs.~\cite{Kaiser_2000, Tzeferacos_2015}) and has been previously used in hybrid (kinetic ion) simulations~\cite{Thoma_2017}.
This type of model has also been extended to incorporate full-wave effects near critical surfaces where the geometric optics break down~\cite{Basko_2017}.

We note that this approach differs from many full EM-wave simulations in laser-plasma interaction, which are most commonly used for direct simulation of laser-plasma instabilities and laser-driven particle acceleration (e.g., Ref.~\cite{Arber_2015}). Models retaining the full EM-wave propagation must resolve the laser wavelength and frequency, which is computationally demanding, so that approach is usually limited to rapid phenomena and small-volume simulations.  PIC models with full EM waves and Monte-Carlo collisions have been developed to study IB \cite{Zhang_2022}, though there are subtleties of the particle collisions during IB absorption \cite{Johnston_1973, Turnbull_2023} which may not be captured in standard PIC Monte-Carlo collision routines \cite{Takizuka_1977}.
At an even more fundamental description, IB has also been simulated with full-wave molecular dynamics simulation \cite{Devriendt_2022}.   This type of model imposes the minimum amount of additional assumptions and, for example, directly resolves all particle collision processes, enabling prediction of transport factors such as the Coulomb logarithm.  However, this method is the most computationally expensive and prohibitive for large-volume HED plasma simulations.  

The model developed here follows the ray-tracing technique with IB absorption.
The primary design choices were motivated by the goal of obtaining a fast, simple model that can be run in-line but with enough physics to compare to radiation-hydrodynamic simulations. This is to allow direct comparisons 
between a full kinetic treatment and radiation-hydrodynamic simulation of laser-heated plasmas.
Rays are followed in 1-D (without refraction in the simulation domain) but with a technique that includes oblique incidence effects and ray reflection.
The model is implemented in the Plasma Simulation Code (PSC) \cite{Germaschewski_2016}. Two companion manuscripts \cite{Lezhnin_2024, Totorica_2024} present benchmarking of the laser deposition and plasma evolution with comparison to radiation-hydrodynamic simulations, thereby presenting additional dynamical verification of this model.

Section \ref{sec:theory_background} reviews the theoretical model for calculating the power deposited by a laser when propagating through a plasma. The numerical implementation is presented in Section \ref{sec:numerical_method}. Additionally, the method for transferring the energy to the particles is described in Section \ref{sec:particle_heating}.
In Section \ref{sec:results}, we present benchmarking against analytical ray-tracing solutions and another ray-tracing code, 
as well as benchmarking of the energy conservation.  We briefly discuss some miscellaneous
implementation notes of the algorithms in a parallel code in Section~\ref{sec:implementation_notes}, and we give our conclusions in Section~\ref{sec:conclusions_and_remarks}.

\section{\label{sec:theory_background}
Laser deposition model}

In this section, we recapitulate the laser propagation 
 and inverse-Bremsstrahlung absorption in a plasma 
to present the model equations 
for ray-tracing and energy deposition.

For a laser propagating through a plasma, it is convenient
to follow the laser intensity along rays rather 
than the full EM field evolution.
This approach is valid when the 
laser wavelength is much shorter than 
other scales in the plasma.  In this case, the 
laser propagation follows from geometric optics,
which involves refraction and reflection 
when propagating through density gradient regions,
and absorption.  
The spatial absorption of the laser energy
due to collisions in the plasma is given by
\begin{equation}
    P_{\text{abs}}= K I_s = - \frac{dI_s}{ds},
    \label{eq:power_heating}
\end{equation}
where $s$ indicates the spatial distance along the ray path.  Here we use $I_s$ as the laser intensity through a 
plane normal to $s$.

Following Dawson, \textit{et al.} \cite{Dawson_1964}, the 
inverse-Bremsstrahlung absorption coefficient $K$ for a plasma is 
\begin{equation}
    K = K_0 \frac{n_e^2}{n_{\rm cr}} \frac{1}{(1 - n_e/n_{\rm cr})^{1/2}},
    \label{eq:K}
\end{equation}
where $K_0$ is the absorption factor, $n_e$ is the local electron density, and $n_{\rm cr}$ is the critical density \cite{Dawson_1964}.  At the critical
density $n_{\rm cr}$, the laser frequency $\omega = 2\pi f$ equals the 
local plasma frequency $\omega_{\rm pe} = (4\pi n_e e^2 / m_e)^{1/2}$.
The absorption factor $K_0$ is 
\begin{equation}
    K_0 = \frac{4}{3} \left ( \frac{2 \pi}{m_e} \right )^{1/2} \frac{Z_{\eff} e^4}{c\, (k_B T_e)^{3/2}} \ln{\Lambda_{\rm IB}},
    \label{eq:K0}
\end{equation}
Here $m_e$ is the electron mass, $Z_{\eff} = (1/n_e) \sum_i n_i Z_i^2$ is the effective ion charge for collisions
summed over ion species $i$, $e$ is the electron charge, $c$ is the speed of light, $k_B$ is the Boltzmann constant, $T_e$ is the electron temperature, and $\ln{\Lambda}_{\rm IB}$ is the Coulomb logarithm for IB \cite{Johnston_1973}.
In conventional units,
\begin{equation}
K = 9.74 \times 10^{-17} \; \frac{n_e^2}{n_{\rm cr} (1-n_e/n_{\rm cr})^{1/2}} \frac{Z_\eff \ln \Lambda_{\rm IB}}{T_e^{3/2}} \; \mathrm{cm}^{-1}
\end{equation}
with densities in cm$^{-3}$ and temperatures in eV.

As described in Johnston, \textit{et al.} \cite{Johnston_1973},
the Coulomb logarithm for a high-frequency laser-plasma interaction,
when $\omega > \omega_{pe}$,
differs from the standard Coulomb log for
low-frequency transport \cite{NRL_formulary}
by the replacement $v_{te}/\omega_{pe} \to v_{te}/\omega$ in evaluating the collisional maximum impact parameter.
Recent work~\cite{Turnbull_2023} highlighted the importance of such corrections, among other atomic physics-related corrections, to correctly predict laser absorption via IB in an experiment.
(To allow flexible comparison against other community codes
and experimental data, in our implementation we keep $\ln{\Lambda}_{\rm IB}$ 
to be a user-specified function.)

Shearer \cite{shearer_1971} provided a useful specialization of the 
laser propagation and deposition problem  for the case of 
oblique laser incidence onto a 1-D plasma density gradient,
which we take as along the $z$ direction.
In this treatment, the laser has a $k$-vector component
both along and perpendicular to the density gradient, 
$\mathbf{k} = k_0 (0, \sin \theta, \cos \theta)$.
We will use this as the basis for our 1-D ray-trace model.

The most important consequence of the 
oblique incidence is that the laser will
 reflect and reverse propagation when it reaches
 a maximum depth of penetration where the density
 reaches
\begin{equation}
    n_m = n_{\rm cr} \cos^2{\theta_0},
    \label{eq:Nm}
\end{equation}
where $\theta_0$ is the laser incidence angle in vacuum.  
If the rays reach a reflection point where
$n_e > n_m$, the ray is reflected to the 
opposite $z$ direction and the ray-trace
continues in the opposite direction back out of the domain.

Next, in the oblique incidence case, we replace the initial intensity $I_{s0}$ with 
\begin{equation}
    I_{z0} = I_{s0} \cos \theta_0,
\end{equation}
to account for the projected intensity on-target.
Finally, we change variables from the propagation direction $s$ to 
the target normal $z$, which gives \cite{shearer_1971}
\begin{equation}
    \frac{d I_z}{dz} = - K_z I_z = -K_0 \frac{n_e^2}{n_{\rm cr} \cos \theta_0}  \frac{1}{(1 - n_e/n_m)^{1/2}} I_z.
    \label{eq:Kz}
\end{equation}
This gives the evolution of the laser intensity and
power deposited along the target normal direction $z$.

This model for laser power deposition into a plasma is convenient because it can be 
directly integrated in 1-D, which, for specified 
laser parameters, plasma density, and temperature,
gives
\begin{equation}
    I_z(z') = I_{z0} \exp\left(- {\int^{z'}_{z_0}{K_z dz}}\right),
    \label{eq:intensity_over_K}
\end{equation}
where $z_0$ is the position from which the
rays are launched.

After obtaining $I_z(z')$, the power deposited
is found from conservation of energy,
\begin{equation}
    P_{\text{abs}}(z') = - \frac{d I_z (z')}{d z}.
    \label{eq:power2}
\end{equation}
In the case of reflection, the power deposited
is summed over the in-going and out-going rays.

This model provides a direct prescription of laser deposition to run in-line in a simulation. At each grid point, the simulation provides the plasma density and temperature input needed to determine $K_z$. The model can be run in higher-dimensions (i.e., 2-D or 3-D); for example, in the case where the laser intensity has a profile as a function of 
transverse position, as in $I_0 = I_0(x_0,y_0)$.  In this case, the solution requires a simple loop over separate $z$ ray-traces for each $(x,y)$. A time-profile of the laser intensity $I_0 = I_0(t)$ can easily be handled by loading a prescribed initial intensity as a function of simulation time.

We note this model includes the global oblique
laser incidence effects, via Eqs.~\ref{eq:Nm}--\ref{eq:Kz},
but does not include further refraction of the laser
in the simulation domain,
as is calculated in more complex ray-tracing
codes \cite{Kaiser_2000}.  
This limit is valid when the plasma
has expanded a distance $L_z$ off the target
which is smaller than the laser diameter, so that
the plasma still appears nearly 1-D to the 
approaching laser.

\section{\label{sec:numerical_method}Numerical implementation}

\subsection{Integration}

To implement the ray-trace model, we integrate
Eq. \ref{eq:intensity_over_K} for each ray
using laser parameters, 
spatial plasma density, and temperature
information from the simulation to 
calculate the absorption rate $K_z$.
Trapezoidal quadrature is performed from 
a specified initial launch point of the rays $z_0$
through the domain until a point of reflection at the modified critical density or a final specified
ray-trace position is reached.  If the ray reflects, the ray is traced back along its path out of the plasma. 
From energy conservation, the power
deposited into the cell is then
determined from
\begin{equation}
    P_{\text abs} = -\frac{\Delta I_z}{\Delta z},
\end{equation}
where $\Delta z$ is the grid spacing.
The power deposited along the inward propagating 
and reflected rays are summed to give a total power deposited at each spatial location along $z$. 

It is straightforward to apply different
laser intensities as a function of time,
since the ray-trace is done for a given
time step of the simulation.  Secondly, when the simulation is run in higher-dimensions,
the user can prescribe a transverse intensity
profile for the laser $I_0(x,y)$, and 
separate ray traces are done along $z$ for each
transverse ray position.
The final result is the power deposited 
$P_{\text abs}(x,y,z,t)$
through the domain at the timestep $t$,
plus auxiliary diagnostic data such as the 
cumulative $K_z$ integral (optical depth) 
along the ingoing and outgoing rays.

\subsection{\label{sec:last_point_calculation}Analytical calculation at the point of reflection}

At the maximum density of penetration ($n_m$), there is a singularity in Eq.~\ref{eq:intensity_over_K} that prevents us from using simple numerical quadrature. The singularity is physically related to the reflection of the laser light. However, the actual energy deposited is integrable, so to obtain an accurate solution for the power deposited near this point, the following is performed. First, the trapezoidal quadrature is stopped at the grid point immediately before the point of reflection, which we denote $z_r$.  Second, an analytical calculation is made for ray-trace and power-deposited between $z_r$ and the physical maximum distance of penetration $z_m$, defined by $n_e(z_m) = n_m$, which may not lie exactly on one of our discrete spatial grid points.  Finally, the numerical integration is continued outward, starting again from $z_r$. 

The intensity of the ingoing laser at $z_r$, denoted $I_{-}$, is 
\begin{equation}
    I_{-} = I_0 \exp\left({- \int_{z0}^{z_r} K_z dz}\right)
\end{equation}
and is first found according to the standard ray-tracing
procedure outlined above.

The intensity of the outgoing laser at $z_r$ after reflection, denoted $I_{+}$, is then
\begin{equation}
    I_{+} = I_{-} \exp\left({- 2 \int_{z_r}^{z_m} K_z dz }\right).
    \label{eq:intensity2}
\end{equation}

\begin{figure}[hbt!]
    \centering
         \centering
         \includegraphics[width=\linewidth]{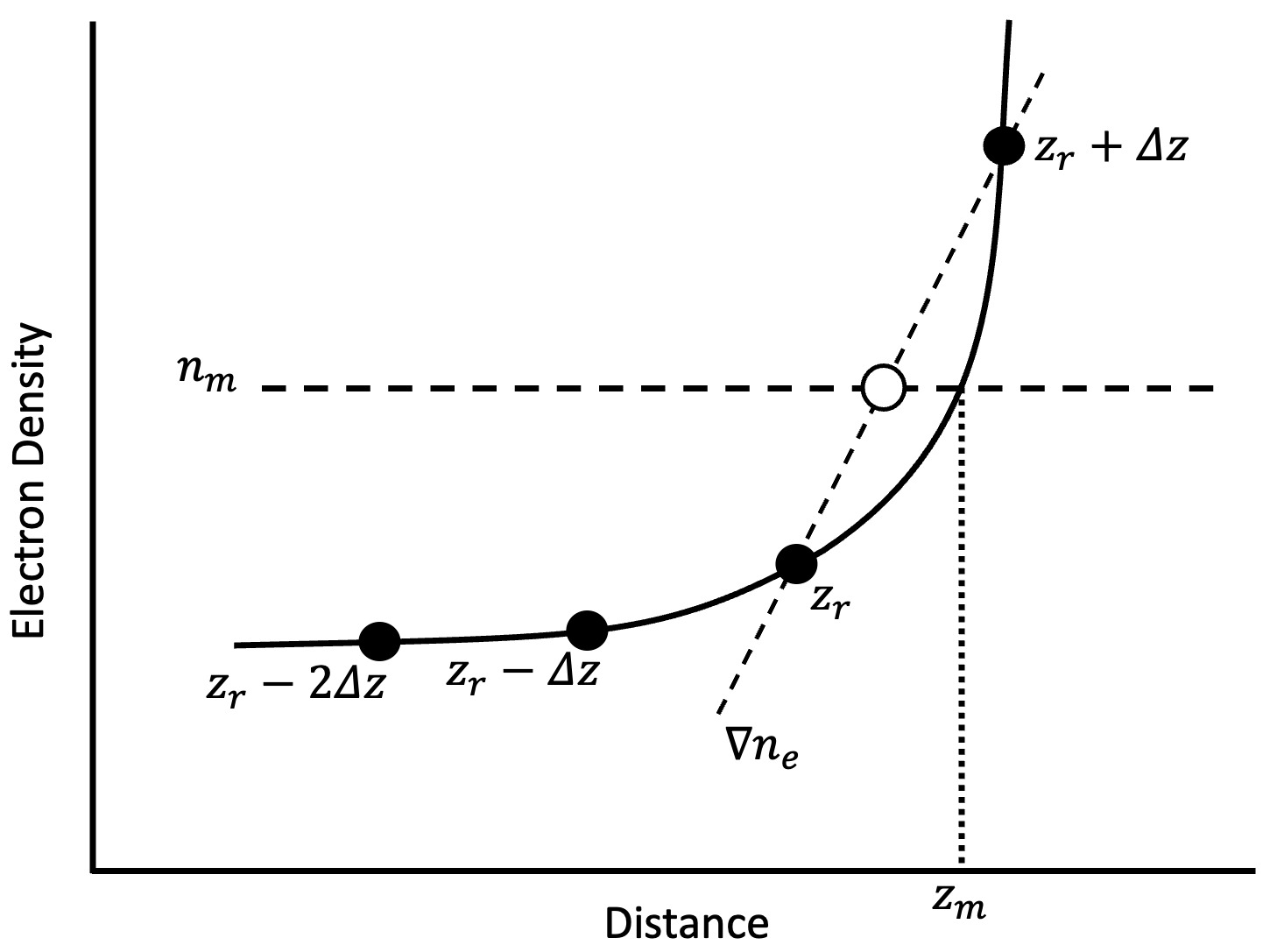}
         \label{subfig:constdens_electron_density}
    \caption{Analytical linear approximation near the turning point. In this example, $z_m$ is not on our spatial grid, $z_r$ is on the spatial grid, and $\Delta z$ is the distance between spatial grid points. $n_m$ is known. The density at the black points is known since they fall on our spatial grid. The location of $z_m$ is unknown and approximated linearly, as depicted by the white point.}
    \label{fig:Linear_Intensity}
\end{figure}

The absorption between $z_r$ and $z_m$ is important as this region often has a significant amount of power deposited resulting from the growth of the denominator in the integral. The solution to calculate the power deposition in this region is to assume a linear density profile between $z_r$ and $z_m$ and use an analytical solution for
the integral to obtain the power deposited between $z_r$ and $z_m$. The density gradient is assumed to be linear
\begin{equation}
    n_{e}(z) \approx n_{z_{r}}+(n_{m}-n_{z_{r}})\frac{z-z_{r}}{\delta},
\end{equation}
where $\delta = z_m - z_r$, and the 
turning point location $z_m$ is inferred from a linear interpolation between
$z_r$ and the next gridpoint.  Figure~\ref{fig:Linear_Intensity} illustrates
the geometry near the turning point.

The analytical solution for $\int_{z_r}^{z_m} K_z dz$ using a linear density gradient is then
\begin{eqnarray}
\int_{z_{r}}^{z_{m}} K_z\, dz =  \frac{2}{15} \frac{\delta\, K_0}{\cos \theta} \frac{ (8 n_{m}^{2}+4 n_{m} n_{z_{r}} + 3 n_{z_{r}}^{2})} {n_{\rm cr}(1-n_{z_{r}}/n_{m})^{1/2}}
\end{eqnarray}

The additional power deposited at the turning point is, accordingly,
\begin{equation}
    \Delta P_{\text{abs}}(z_r) = \frac{I_{-} - I_{+}}{\Delta z}.
\end{equation}

\subsection{\label{sec:particle_heating}Particle Heating}

From the above development, we have the power deposited per unit volume at all points along the path of the laser. 
Within a particle-in-cell code, each macro-particle 
represents a weighted contribution to the total electron density $n_e$.
Dividing $P_{\text{abs}}(x,y,z)$ by the local density $n_e(x,y,z)$ gives the power deposited by the laser per electron.  The heating module then performs a 
randomized kick to each macroparticle in the 
associated cell such that the 
particle's energy will increase on average by
\begin{equation}
    H(x,y,z) = \frac{P_{\text{abs}}(x,y,z)}{n_e(x,y,z)} \Delta t_{\rm heating},
    \label{eq:energy_per_particle}
\end{equation}
where $H$ is the energy deposited per electron,
and $\Delta t_{\rm heating}$ is the time-step
between calls to the heating module.
Per the standard IB theory, given that $m_e \ll m_i$,
the heating is applied only to the electrons,
which subsequently can heat the ions via
standard Coulomb collisions.
The energy is distributed to the electrons in a statistical
manner by giving each particle an independent randomized momentum kick defined by
\begin{equation}
\Delta {\bf p} = \sqrt{ \frac{2}{3} m_e H} \; \left ( r_{1} \hat{\bf{x}}+r_{2} \hat{\bf{y}} + r_{3} \hat{\bf{z}}\right),
\end{equation}
where the numbers $r_{1}$, $r_{2}$, and $r_{3}$ are randomly sampled 
from a standard normal distribution. The average energy change per particle is
then given by
\begin{equation}
    \begin{aligned}
\langle \Delta \epsilon \rangle &= \langle ({\bf p} + \Delta {\bf p})^{2}/2m_e -{\bf p}^{2}/2m_e \rangle \\ 
&= \langle 2({\bf p}\cdot \Delta {\bf p})\rangle /2m_e +\langle \Delta {\bf p}^{2}\rangle/2m_e.
    \end{aligned}
\end{equation}
Assuming a well-sampled distribution of particles 
and small-amplitude kicks uncorrelated to the particle momenta, $\langle ({\bf p}\cdot \Delta {\bf p})\rangle \approx 0$, and the chosen amplitude, leads to the desired average energy change per particle
\begin{equation}
    \begin{aligned}
    \langle \Delta \epsilon \rangle \approx (2/3)m_e H \left (\langle r_{1}^{2}\rangle+\langle r_{2}^{2} \rangle +\langle r_{3}^{2}\rangle \right ) /2m_e = H.
    \end{aligned}
\end{equation}

We note that for the time being, we have implemented the simplest possible kicking mechanism, which is to heat the plasma via Gaussian kicks,
drawn from a uniform distribution for all particles.  However, more advanced Langevin schemes have been developed that kick the particles in a fashion relevant to IB heating \cite{Detering_2002}, which predominantly heats the coldest particles in the distribution, leading to super-Gaussian particle distributions and the well-known Langdon effect \cite{Langdon_1980}. Implementing such advanced kicking schemes will be pursued in future implementations.

\section{\label{sec:results}Results and Benchmarks}

In this section, we verify the ray-tracing module via several benchmarks.  
First, we verify that the code and quadrature scheme obtains the expected absorption for analytic plasma profiles, and second, that the scheme conserves energy between the input laser energy and changes to the plasma energy. Finally, we cross-benchmark the ray trace module against a 2-D ray-tracing calculation in the
radiation hydrodynamics code FLASH, which includes in-plane laser refraction, and thereby validates how
our reduced 1-D model treats oblique incidence.
Two separate papers show additional dynamic benchmarking with ablation of a plasma slab and comparison to various radiation hydrodynamics simulations, verifying along the way that the laser deposition is well-matched between the simulations \cite{Totorica_2024, Lezhnin_2024}.

\begin{figure}[hbt!]
    \centering
         \centering
         {\includegraphics[width=\linewidth]{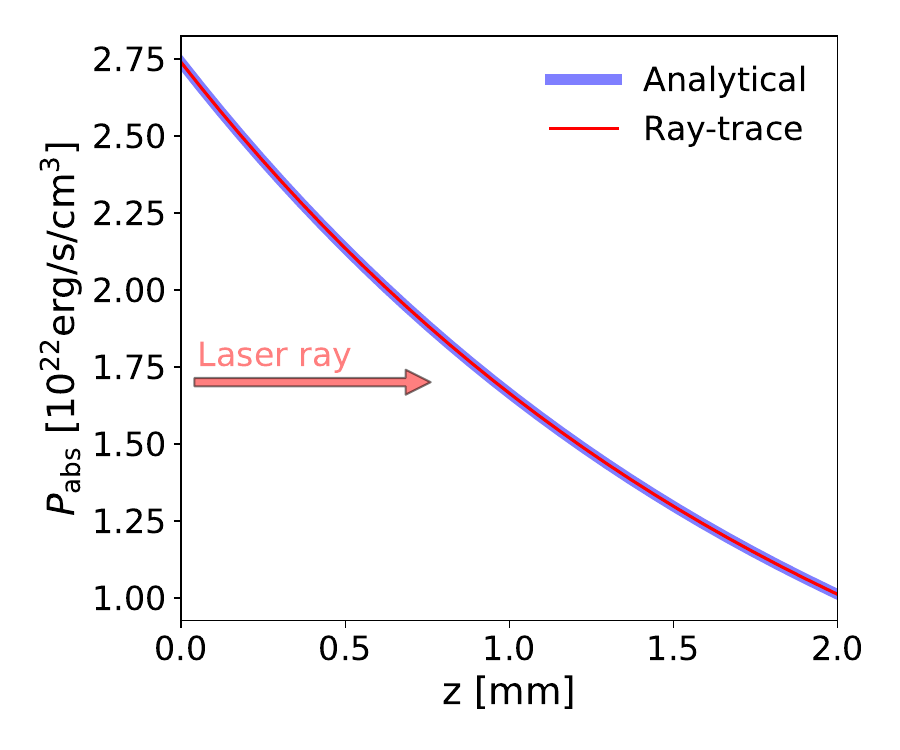}}
         \label{subfig:constdens_power}
     
    \caption{Analytical and numerical laser power deposition $P_{\rm abs}$ for a constant density profile. The laser at wavelength 351~nm, with intensity $5.5 \times 10^{21}$~erg/cm$^2$/s, enters from the left and propagates towards the right. The laser is normally incident ($\cos \theta_0 = 1$) onto the uniform plasma with an electron density of $10^{21}\,\rm cm^{-3}$, effective charge state $Z_{\eff}=5.29$, and electron temperature of $1800$ eV.  The results of the numerical and analytical solutions are identical.}
    \label{fig:constdens}
\end{figure}

\begin{figure}[hbt!]
    \centering
         \centering
         {\includegraphics[width=\linewidth]{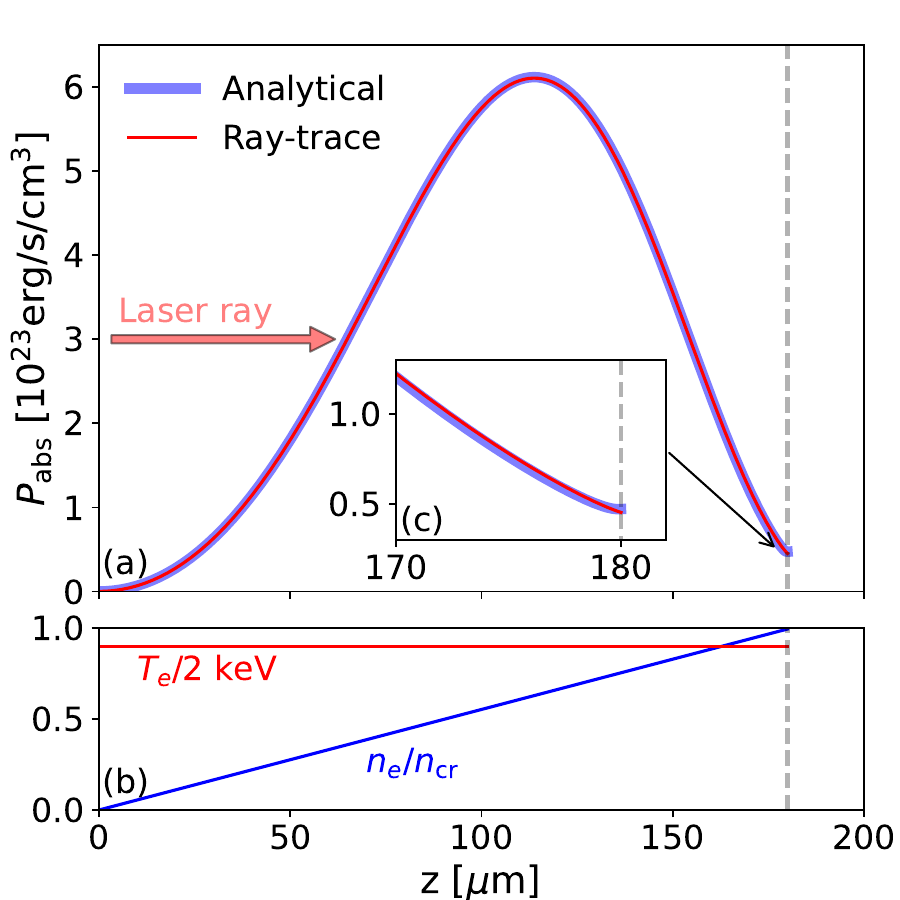}}
         \label{subfig:lineardens_power_close}
     
    \caption{(a) Laser power absorption in a linear electron density profile calculated analytically (blue) and with the ray-trace model (red). The laser enters from the left and propagates towards the right at normal incidence. (b) Electron density and temperature profiles. (c) A close-up view near the point of reflection where our analytical technique is used for the numerical solution. The point of reflection is shown by the vertical dashed line.}
    \label{fig:lineardens}
\end{figure}

Figures \ref{fig:constdens} and \ref{fig:lineardens} utilize simple electron density profiles ($n_e$) to benchmark the validity of the 1-D ray trace and numerical integration.   These benchmarks test the numerical ray trace for a static plasma.
The electron density profiles used for Figs.~\ref{fig:constdens}a and \ref{fig:lineardens}a are analytically solvable to provide the laser intensity at each spatial location along the path of the laser.

The first benchmark for our laser-energy deposition model is the case of a constant-density plasma with a sub-critical density under the maximum density of penetration ($n_e < n_m$).  At a density under $n_m$, the laser propagates through the plasma without reflection. In this constant density case, the analytical solution for the intensity of the laser is
\begin{equation}
I (z') = I_0 \exp \left( - \frac{K_0}{n_{\rm cr} \cos{\theta_0}} \frac{n_e^2}{(1 - n_e/n_m)^{1/2}} (z' - z_0)\right),
\label{eq:analytical_const_sol}
\end{equation}
where $n_e$ is a constant and $(z' - z_0)$ is the 
distance of laser propagation.

Figure \ref{fig:constdens}, shows the result from the numerical raytrace calculation matches exactly with that of the analytical solution.  The laser, at wavelength 351~nm with intensity $5.5 \times 10^{21}$~erg/cm$^2$/s, enters from the left and propagates towards the right. 
For this laser wavelength, the critical density ($n_{\rm cr}$) is $9.0496\times10^{21}\, \text{cm}^{-3}$.
The laser is normally incident ($\cos \theta_0 = 1$) onto the uniform plasma of electron density of $10^{21} \rm cm^{-3}\approx 0.11 n_{\rm cr}$, effective charge state $Z_{\rm eff}=5.29$, and electron temperature of $1800$ eV.
The figure shows excellent agreement between the
ray-trace model and analytical calculation.

A second benchmark is a linear electron density profile with a peak density higher than $n_m$, which was originally calculated analytically in Ref.~\cite{shearer_1971}.  
This benchmark has the benefit of testing the calculation at and near the point of reflection as well as benchmarking the backward propagating ray after reflection. In the case of a linear density gradient, the density profile follows
\begin{equation}
    n_e = n_0 \frac{z}{L}.
    \label{eq:lin_density}
\end{equation}
For the test shown in Fig.~\ref{fig:lineardens}, we use $n_0 = 10^{22}$~cm$^{-3}$, $L$ = 200~$\mu$m, and the same laser parameters as for the constant density case. The critical density is again
$\approx 9 \times  10^{21}$~cm$^{-3}$, which is reached
at $z_m \approx 180~\mu$m.  The analytical and numerical solutions are shown to be in overall excellent agreement.

\begin{figure}[h!]
\begin{center}
\includegraphics[width=\linewidth]{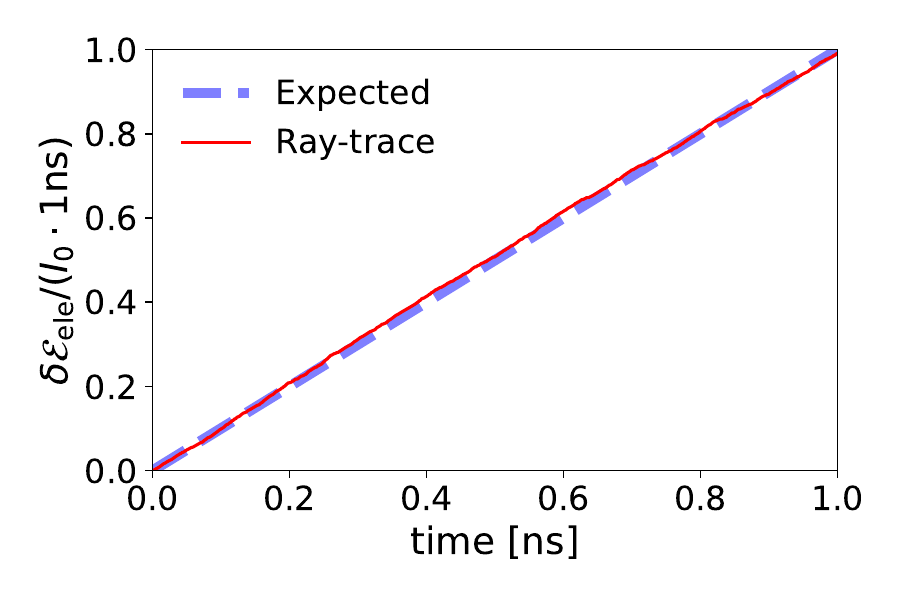}
\caption{\label{fig:energy_conservation} 
Temporal evolution of the internal energy areal density gain of the simulated electrons, $\delta \mathcal{E}_{\rm ele}$ (red solid line), compared to the theoretical prediction for full absorption of a laser with intensity $I_{0} = 8 \times 10^{16}\;\mathrm{erg/cm^{2}/s}$, $I_0 t$ (blue dashed line).
}
\end{center}
\end{figure}

As the next benchmarking test, we confirm the overall energy conservation between laser energy input and plasma heat by simulating the irradiation of a uniform electron-ion
plasma and comparing the change in the total energy of the plasma 
with the expected laser absorption.  For this test, we assume 
$T_{e} = T_{i} = 10\;\mathrm{eV}$ and $n_{e} = 10^{20}\;\mathrm{cm}^{-3}$ for the plasma,
$\lambda = 1.064\;\mu\mathrm{m}$ for the laser wavelength,
and $I_{0} = 8 \times 10^{16}\;\mathrm{erg/cm^{2}/s}$ for the laser
intensity. The one-dimensional simulation domain has a length of
$\approx 10\;\mathrm{mm}$ and is resolved by a spatial grid with 
2000 grid cells and 100 particles per cell for each plasma 
species.  The simulation is run for $\approx 63000$ timesteps, 
corresponding to a total duration of $1\;\mathrm{ns}$. The 
particle boundaries are reflecting, and the electromagnetic field solver is 
turned off to isolate the energy change from the laser module.  For these parameters, the laser is fully absorbed (to
machine precision) approximately midway through the simulation 
domain, penetrating slightly deeper into the plasma with time as
the electron temperature increases and the absorption decreases 
near the laser input boundary of the domain. Figure 
\ref{fig:energy_conservation} shows the temporal evolution of the
total energy of the simulated electrons calculated every 100 
timesteps (red), which is seen to be in close agreement with the 
theoretical prediction for full absorption of the laser (dashed 
line).

\begin{figure}
    \centering
    \includegraphics[width=\linewidth]{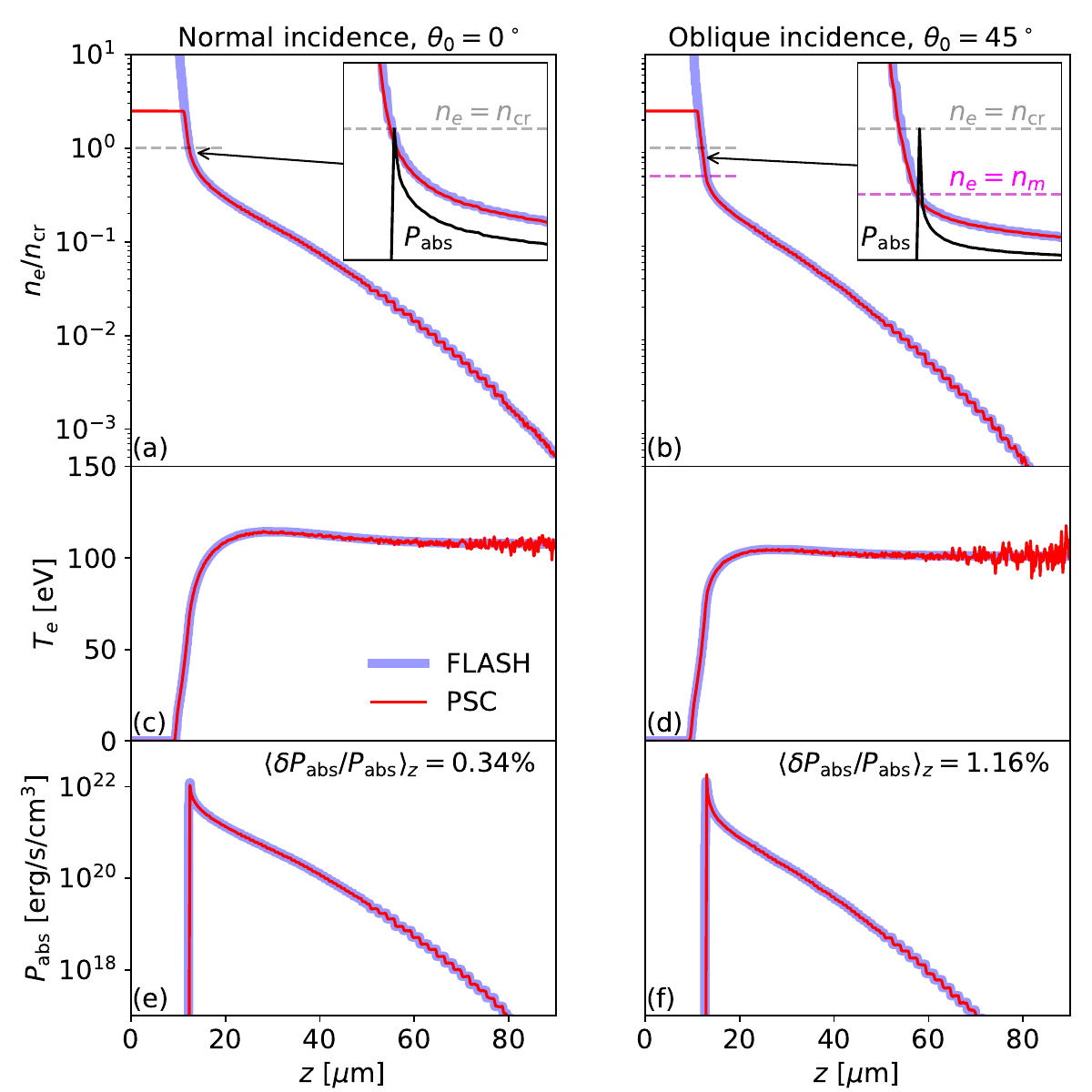}
    \caption{Testing laser ray tracing and laser energy deposition module in PSC against the FLASH code. (a,b) Electron density, (c,d) temperature, and (e,f) laser power absorption for (a,c,e) normal and (b,d,f) oblique incidence cases. Insets in (a) and (b) depict the zoom-ins around $z_r$ and $z_m$, respectively. Very good agreement between PSC and FLASH simulations is seen.}
    \label{fig:PSCvsFLASH}
\end{figure}

Finally, we cross-benchmarked the present laser-ray trace module against the FLASH radiation hydrodynamics code~\cite{Tzeferacos_2015}.   
Notably, the FLASH simulations used a 2-D domain, where a propagating laser can refract within the simulation plane. This allows us to test the present 1-D model, with the oblique incidence corrections of Eqs.~\ref{eq:Nm}-\ref{eq:Kz}, against a more fundamental model. (Additional benchmarking, including the cross-benchmarking of the subsequent dynamical evolution against FLASH, is described in detail in a companion manuscript~\cite{Lezhnin_2024}.) The setup of the test is as follows.  We started with the LaserSlab test problem provided with the FLASH distribution~\cite{FLASH_manual}.  We conducted 2-D FLASH simulations of laser ablation of a solid aluminum slab with initial density of 2.7 g/cm$^3$ and a 50 $\mu$m thickness by a laser pulse of 0.3 ns duration with a 0.1 ns linear rise to a flat-top peak intensity of $3.54\times10^{11}\, \rm W/cm^2$, $1\,\mu \rm m$ wavelength, and a uniform transverse profile.  One of the simulations had normal incidence ($\theta_0 = 0^\circ$) and another used oblique incidence ($\theta_0 = 45^\circ$).  The simulation box was 400 by 400 microns, divided into 2$\times$2 blocks containing 16$\times$16 cells each, with a maximum level of adaptive mesh refinement set to 7. 
We simulated the plasma evolution up to 0.3 ns and then took a cut along the target normal to obtain 1-D profiles for both cases. Then, we conducted two short 1-D PSC simulations initialized directly from these FLASH profiles, running a small number of steps until the first call to the laser ray-trace and deposition module. 
We adopted a Coulomb logarithm in PSC that identically matched the default version in FLASH.  Figure~\ref{fig:PSCvsFLASH} compares PSC and FLASH in terms of the electron density (a,b), temperature (c,d), and laser power deposition (e,f) profiles for normal incidence (a,c,e) and oblique incidence (b,d,f) cases.  
Note that while the density can be directly loaded into PSC, the temperature will obtain statistical fluctuations due to the 
finite number of computational particles (Fig.~\ref{fig:PSCvsFLASH}c,d).
Nevertheless, for these matched density and temperature profiles, the
 resulting laser absorption profiles are in excellent agreement, as shown in Fig.~\ref{fig:PSCvsFLASH}e,f. 
Insets in Figs.~\ref{fig:PSCvsFLASH}a,b depict a zoom-in onto the laser ray reflection region. In the normal incidence case, we see that both FLASH and PSC capture the critical surface location, $z_{r}$, observable as the location where the laser absorption profile reaches a maximum and then cuts off in Fig.~\ref{fig:PSCvsFLASH}e.  In the oblique incidence case, we see that both FLASH and PSC capture an identical maximum penetration depth $z_{m}$, now given by $n_e(z_m) = n_m =  n_{\rm cr} \cos^2\theta_0$. The agreement between the full ray-tracing approach utilized in FLASH and our model using the oblique incidence effects of Ref.~\cite{shearer_1971} demonstrates that this model is
effective for capturing the global oblique incidence effects on laser absorption in realistic scenarios.

\section{\label{sec:implementation_notes}Implementation notes}

The ray-trace routine was implemented as a 
separate module in PSC.  The present version 
allows transverse laser intensity profiles, with
the restriction that the ray-tracing
itself is along one
coordinate axis.  In this case, a bundle of 
rays is ray-traced, with one ray per transverse cell
in $x$ and $y$, 
with each ray's initial intensity set by
the intensity profile $I_0(x,y)$.

PSC is implemented using MPI with
domain decomposition, where 
each processor holds the electromagnetic fields
and particles associated with particular spatial tiles 
of the global domain \cite{Germaschewski_2016}.  Periodically, the 
particle data is used to calculate the 
moments (density, flows, and temperatures), again 
held on each processor, which is used
for both output and now the ray-trace
model.

The ray-trace requires a spatial integral over large
volumes of the domain, much larger
than the subdomain assigned to each processor. 
In the present
version, the relevant moment data (density $n_e$, 
electron temperature $T_e$), 
needed for calculating the terms in Eq.~\ref{eq:K},
are gathered to one processor (via \texttt{MPI\_reduce}),
which then performs the ray-trace 
and calculates the laser-energy deposition.
The laser-energy deposition information 
is then distributed
back to the full 
set of processors 
(via \texttt{MPI\_scatter}), after which each
processor kicks the particles in its own
subdomain.

We found that the global data gather 
could be memory-restrictive for large 2-D simulations.
A first optimization, which we have implemented, is to
restrict the 
ray-tracing to a finite subdomain, which limits
the memory and communication requirements.  This
is appropriate when considering plasma ablation
from solid targets, where much of the domain can
remain at very low density and therefore negligible absorption.
A second optimization, planned for the future, will be 
to parallelize the ray-trace using several stripes 
oriented along the ray-tracing direction with
a larger set of ray-tracing processors; such a 
data subdivision is facilitated by the
fact that the ray-tracing is only along $z$.

We found the ray-trace and particle kicks can be significantly
sub-cycled compared to the PIC time-step,
since the ray-trace will only yield different 
results after characteristic dynamic times $(\sim L/V)$,
where $L$ is the system size and $V$ is a typical
plasma velocity, whereas the PIC timestep
is required to be on the order of the plasma frequency, $\sim \omega_{pe}^{-1}$,
with $\omega_{pe}L/V \gg 1$.   Sub-cycling is important because
the ray-tracing requires a global communication.  Furthermore,
the ray-trace also requires a recent calculation of the 
particle moments (density, temperature), which requires a computationally-expensive
iteration over the full particle array (by each processor);
however, the moments are usually already available since they are periodically 
calculated for simulation output.
Second,
the particle kicks can also be sub-cycled as long as 
$H / T_e \ll 1$, where $H$ is the magnitude
of the kick given to the particles; this is useful because 
again the particle kick step requires an iteration through the full 
particle array.

\section{\label{sec:conclusions_and_remarks}Discussion and Conclusions}

The present paper describes a ray-tracing model for laser-energy deposition, which
has been implemented inline in a plasma particle-in-cell code. This will allow kinetic simulation of laser-solid interaction and the physics of ablated plasmas for high-energy-density physics and laboratory plasma astrophysics experiments \cite{Fox_2013,
Schaeffer_2020, Fox_2018, Fox_2011, Lezhnin_2021}.   The model uses a reduced 1-D ray-trace that, however, includes oblique incidence effects, which are important for modifying plasma absorption. The model was benchmarked against analytic solutions 
and against a 2-D laser ray-tracing code with excellent agreement.  Separate papers have presented additional dynamic benchmarking of the laser ray-trace module against plasma evolution in radiation hydrodynamics simulations and have begun to show the kinetic plasma physics, which can now be studied with this multi-physics capability \cite{Totorica_2024, Lezhnin_2024}. 

\bibliography{raytrace}

\end{document}